\documentclass[aip,reprint]{revtex4-2}

\usepackage{graphicx}
\usepackage{amsmath,amssymb}
\usepackage{hyperref,url}
\usepackage{siunitx}
\usepackage{mathcomp}
\usepackage{xcolor}
\newcommand{\Ma}{\mathrm{Ma}}
\newcommand{\Ev}{\mathrm{Ev}}
\newcommand{\xiavg}{\xi_\text{avg}}
\newcommand{\figref}[1]{Fig.~\ref{#1}}

\newcommand{\reviewerA}[1]{#1}
\newcommand{\reviewerB}[1]{#1}
\newcommand{\reviewerAB}[1]{#1}
\begin{document}


\title{Non-monotonic surface tension leads to spontaneous symmetry breaking in a binary evaporating drop} 



\author{Christian Diddens}
\email[]{c.diddens@utwente.nl}
\affiliation{Physics of Fluids group, Department of Science and Technology, Max Planck Center for Complex Fluid Dynamics and 
J. M. Burgers Centre for Fluid Dynamics, University of Twente, P.O. Box 217, 7500 AE Enschede, The Netherlands}
\author{Pim J. Dekker}
\affiliation{Physics of Fluids group, Department of Science and Technology, Max Planck Center for Complex Fluid Dynamics and 
J. M. Burgers Centre for Fluid Dynamics, University of Twente, P.O. Box 217, 7500 AE Enschede, The Netherlands}
\author{Detlef Lohse}
\email[]{d.lohse@utwente.nl}
\affiliation{Physics of Fluids group, Department of Science and Technology, Max Planck Center for Complex Fluid Dynamics and 
J. M. Burgers Centre for Fluid Dynamics, University of Twente, P.O. Box 217, 7500 AE Enschede, The Netherlands}
\affiliation{Max Planck Institute for Dynamics and Self-Organization, Am Fassberg 17, 37077 G\"ottingen, Germany}

\date{\today}

\begin{abstract}
The evaporation of water/1,2-hexanediol binary drops shows remarkable segregation dynamics, with hexanediol-rich spots forming at the rim, thus breaking axisymmetry. While the segregation of hexanediol near the rim can be attributed to the preferential evaporation of water, the symmetry-breaking and spot formation could not yet be successfully explained. With three-dimensional simulations and azimuthal stability analysis of a minimal model, we investigate the flow and composition in the drop. We show that a slightly non-monotonic surface tension causes the emergence of a counter-rotating Marangoni vortex in the hexanediol-rich rim region, which subsequently becomes azimuthally unstable and forms the observed spots. \reviewerAB{Accurate measurements with several different methods} reveal that the surface tension is indeed non-monotonic. This work provides valuable insight for applications like inkjet printing or spray cooling. 
\end{abstract}

\maketitle 

Since the discovery of the \textsl{coffee stain effect} by Deegan \textsl{et al.} \cite{Deegan1997} in 1997, the evaporation of sessile droplets has attracted tremendous interest. While the evaporation of isolated single-component droplets has been mainly understood (cf. recent reviews\cite{Gelderblom2022,Wilson2023}), the complexity dramatically increases when multi-component droplets, i.e. droplets consisting of a mixture of components with different volatitilies, evaporate. Thereby, compositional gradients arise in these droplets and due to various physicochemical interactions, multi-component droplets show a plethora of intriguing phenomena\cite{Lohse2020,Wang2022}, e.g. evaporation-induced phase separation\cite{Tan2016}, axisymmetric\cite{ThayyilRaju2022} and symmetry-breaking\reviewerA{\cite{Rowan2000,Christy2011,Bennacer2014,Kim2016,Diddens2017}} solutal Marangoni flow, Marangoni contraction\cite{Karpitschka2017} and shape deformation\cite{Pahlavan2021}, and potential dominance of natural convection over solutal Marangoni flow\cite{Edwards2018,Li2019,Diddens2021}.
Since applications such as e.g. inkjet printing\cite{Lohse2022,Hoath2016} and nano-scale self-assembly\cite{Brinker1999} are based on the evaporation of multi-component droplets, a detailed understanding of all aspects of multi-component droplet evaporation is of utmost relevance.

One particular intriguing and relevant example is the evaporation of droplets consisting of the binary mixture of water and 1,2-hexanediol \reviewerA{(hexanediol in the following)}, which constitutes a simplified model ink for water-based inks. Here, hexanediol is added as a spreading agent, since minor amounts of hexanediol strongly reduce the surface tension and thereby aid to fully cover the paper with a minimal volume of ink.
\begin{figure}[ht!]
\includegraphics[width=8.6cm]{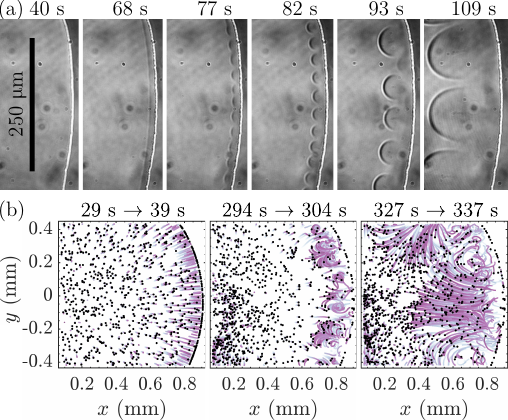}%
\caption{Evaporation of a water + 10 wt$\%$ hexanediol droplet. (a) Transmission detection with a confocal microscope, i.e. with coherent light. \reviewerB{Initial conditions:} $V_0 = 0.25 \, \tcmu$l, $\theta_0 = 12\,\mathrm{^\circ}$,  $T = 23 \,\tccelsius$, $\mathrm{RH} = 30\%$. 
Strong gradients in the concentration are visible due to the difference in refractive index of water $n_\mathrm{w}= 1.33$ and hexanediol $n_\mathrm{hd}=1.44$ \cite{crc2016}.
68 s after the drop has been deposited a sharp gradient forms close to the \reviewerB{rim, which} breaks up at 77 s into small blobs \reviewerB{that} start to coalesce. (b) By adding fluorescent particles (\reviewerB{diameter} $1.1 \, \tcmu$m) and imaging them with the confocal microscope, the flow can be measured close to the substrate, using particle tracking velocimetry. \reviewerB{Initial conditions:} $V_0 = 0.5 \,\tcmu$l, $\theta_0 = 45\,\mathrm{^\circ}$,  $T = 25 \,\tccelsius$, $\mathrm{RH} = 50\%$. See also the Supplementary Movie\cite{supp} \reviewerA{and the Gallery of Fluid Motion\cite{GFM}}. \reviewerA{When inhibiting the evaporation after the onset of spot formation, the spots vanish immediately\cite{SI}.} \label{fig1}}%
\end{figure}
Despite water and hexanediol being miscible in any ratio, these droplets show remarkable segregation dynamics during evaporation. As reported in Refs.\cite{Li2018,Li2020,Li2022} and depicted in \figref{fig1}, the droplet is initially rather homogeneous, but with ongoing time, first a ring forms along the contact line which successively breaks up into individual spots that grow and merge over time.  By the addition of dyes, it has been revealed that the ring and the subsequent spots mainly consist of the nonvolatile hexanediol, whereas the central region and the pathways between the spots contain a considerable amount of water\cite{Li2018}. Remarkably, the formation of these spots is accompanied by strong non-axisymmetric flows, cf. \figref{fig1}(b). Initially, particle tracking velocimetry (PTV) reveals axisymmetric and radially outward flow in the plane close and parallel to the substrate, but the moment the spots form, flow reversal occurs and an arch-like flow pattern emerges within each spot. The entire process shows a striking similarity with the observed dynamics of a water-ethanol mixture evaporating from a Hele-Shaw cell \cite{LopezDeLaCruz2021}.

\begin{figure}[tb]
\includegraphics[width=8.6cm]{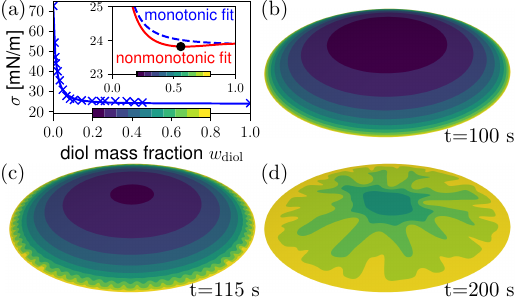}%
\caption{(a) Surface tension data from literature\cite{Romero2007fpe} and a zoomed view on two different fits (the minimum is indicated by the black dot). 3d simulations with the monotonic fit remain axisymmetric (see Supplementary Movie\cite{supp}). With a slightly non-monotonic fit ($<\SI{0.25}{\milli\newton/\meter}$ increase in surface tension after the initial strong decrease), simulations can reproduce the experimental observations for the non-monotonic case. (b-d) results of the 3d simulation at different times. \label{fig2}} 
\end{figure}

Since the spots form after the strong decline in the surface tension (see \figref{fig2}(a)), i.e. in a region where the surface tension is nearly constant, it was concluded that the segregation is a consequence of Marangoni flow cessation\cite{Li2018}. In the absence of recirculating flow, the remaining water can evaporate from the rim region, effectively leading to strong segregation between both constituents. While this explains the high diol concentration at the rim, it fails to provide a mechanism for the axisymmetry breaking, i.e. the spot formation, nor does it explain the remarkably strong flow in the spots. Further studies on water/hexanediol droplets also could not clarify the origin of the emerging hexanediol-rich spots at the rim\cite{Li2020,Li2022}. Other binary mixtures can show segregation by evaporation but without any spot formation\cite{Kim2018}.

\reviewerA{Since hexanediol is 4.7$\%$ less dense than water\cite{Romero2007jct}},
natural convection could break the axial symmetry via a Rayleigh-Taylor instability. While the relevance of density-driven flow in binary droplets can be remarkable\cite{Edwards2018,Li2019,Diddens2021}, a Rayleigh-Taylor instability can be ruled out since the observed process remains the same when reversing the direction of gravity, i.e. by considering pendant droplets instead of sessile droplets. A notable difference in the dynamics between sessile and pendant droplets, i.e. the presence of a Rayleigh-Taylor instability, can only be seen when adding silicone oil as a third component to the droplet\cite{Li2020}. Moreover, 3D simulations of the binary hexanediol/water droplet with a monotonically decreasing surface tension remain axisymmetric, irrespectively of the direction of gravity.

\reviewerA{In principle, phase separation, as frequently seen in ternary systems\cite{Tan2016,Othman2023}, could be a reason for the emerging spots. However, 1,2-hexanediol and water are miscible at arbitrary ratios at room temperature\cite{Li2018}.} Beyond that, phase separation alone, which is driven by (anti-)diffusive fluxes, cannot explain the observed strong recirculating flow in and around the growing spots. \reviewerAB{While the phenomenon appears on different substrates with different contact angles, it is extremely sensitive to impurities, since the deposition of the droplet with a PDMS-coated plastic syringe actually suppresses the observed phenomenon and triggers a Rayleigh-Taylor instability instead\cite{Li2020}}. \reviewerA{In summary, the origin of the symmetry breaking and the spot formation hitherto remained elusive.}

The only remaining potential mechanism which can drive this type of flow is the Marangoni effect, although the monotonic surface tension data of Ref. \cite{Romero2007fpe} suggests that the surface tension quickly attains a constant plateau when the overall fraction of hexanediol is increased (\figref{fig2}(a)). Comparison between experiments and simulations reveal that the spot formation emerges in the plateau region and therefore Marangoni flow was ruled out so far. However, the observation of a flow reversal at the rim, just before the spots start to form, indicates that the surface tension must start to increase again, i.e. it must pass a minimum and in total show a non-monotonic profile \reviewerA{as function of the compositon, whereas thermal Marangoni flow cannot be responsible for this reversal\cite{SI}}. In this scenario, the emergence of the counter-rotating vortex near the rim happens once the segregated hexanediol concentration passes the minimum of the surface tension. \reviewerA{Such non-monotonic surface tension profiles are rare, but can be found e.g. in liquid metal alloys\cite{March2008} and in aqueous 1,5-pentanol solutions\cite{Glinski2000}.}
While evaporation of a binary droplet with a non-monotonic surface tension has not been investigated, a non-monotonic evolution of the surface tension can be achieved rather easily in ternary droplets, where it leads to axisymmetry breaking during spreading\cite{Baumgartner2022}.

To test our hypothesis of a non-monotonic surface tension, we performed 3D simulations of evaporating hexanediol/water binary droplets, analogous to the axisymmetric simulations of Ref.\cite{Li2018} (see Refs.\cite{Li2018,Diddens2017} and the supplementary information of Ref.\cite{Jalaal2022} for the employed finite element method\reviewerB{\cite{pyoomph}}). We used two different fits of the surface tension data from literature\cite{Romero2007fpe}. With the monotonically decreasing fit used hitherto, the droplet remains axisymmetric, but if the fit has a very shallow minimum -- within the experimental uncertainty\cite{Romero2007fpe} of \SI{0.3}{\milli\newton/\meter} -- indeed the same behavior as in the experiments can be reproduced, cf. \figref{fig2}. The flow inside the droplet is initially axisymmetric, driven by the steep decline of the surface tension (\figref{fig2}(b)). When the composition near the rim crosses the surface tension minimum, the Marangoni flow locally reverses. Shortly after this reversal, the simulations show exactly the same features as in the experiments, i.e. the rim region breaks up into individual spots (\figref{fig2}(c)), which subsequently grow and coalesce with ongoing time (\figref{fig2}(d)). 

\begin{figure}[tb]%
\includegraphics[width=8.6cm]{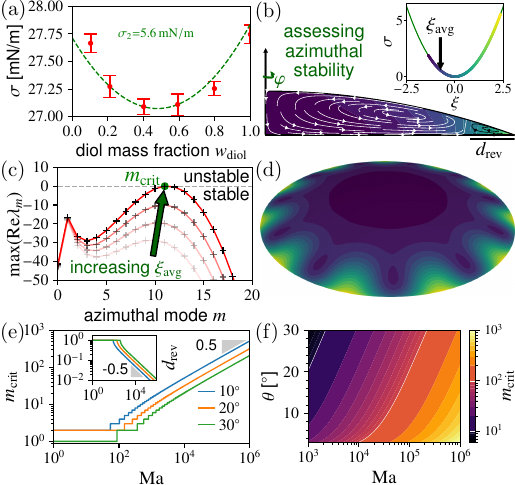}%
\caption{(a) Accurately measured data\cite{PRFJointSubmission} indeed reveal a minimum in the surface tension, which can be locally approximated by a parabola with curvature $\sigma_2$ (dashed line). (b) Axisymmetric base state of the minimal model at criticality (\reviewerA{here for} $\Ma=1500$, $\theta=\SI{20}{\degree}$). (c) Azimuthal bifurcation upon increasing the averaged nondimensional diol concentration $\xiavg$. (d) Plot of the base state plus the critical eigenfunction. (e) Critical mode number as a function of the Marangoni number $\Ma$ and contact angle $\theta$, while the inset shows the size of the flow reversal region $d_\text{rev}$. (f) Same as (e), but for the full range of considered contact angles. \label{fig3}}%
\end{figure}

Since these simulation results provide a good indication that our hypothesis of a slightly non-monotonic surface tension is valid, \reviewerAB{we accurately measured the surface tension via the Du Noüy ring method and a sophisticated pendant drop method\cite{PRFJointSubmission}}. In the surface tension data from literature\cite{Romero2007fpe}, the overall curve appears to be monotonic with a steep decrease followed by a nearly constant plateau. \reviewerAB{In contrast, all our measurements\cite{PRFJointSubmission} could indeed reveal the presence of a very shallow minimum of around $\sim \SIrange[range-phrase=-,range-units=single]{0.3}{0.7}{\milli\newton/\meter}$ in magnitude (\figref{fig3}(a))}. Thereby, we can conclude that the spot formation is caused by first segregation of hexanediol due to the preferential evaporation of water near the rim, followed by a flow reversal due to the non-monotonic surface tension of water/hexanediol binary mixtures and subsequent symmetry-breaking.

To understand why a local minimum in the surface tension gives rise to axial symmetry breaking and the formation of segregated spots, and to quantify the number $m_\text{crit}$ of initially emerging spots, we developed a minimal model. In the spirit of Occam, we only account for the essence, i.e. the non-monotonic surface tension profile, while vastly simplifying the complicated problem in all other aspects. 
For any binary mixture possessing a local minimum in the surface tension curve at the mass fraction $w_\text{min}$, one can locally approximate the surface tension by a parabola, \reviewerA{$\sigma=\sigma_0+\sigma_2(w-w_\text{min})^2/2$} (dashed line in \figref{fig3}(a)). Assuming that the circulation velocity due to Marangoni flow is considerably faster than the motion of the interface and that $\sigma_0$ is sufficiently large to ensure a spherical cap shape despite deformations due to the Marangoni flow, one can aim for quasi-stationary solutions on a static mesh resembling a spherical cap shape. After nondimensionalization with the diffusive time and velocity scales and the contact line radius $R$ one obtains in the isothermal, non-buoyant Stokes limit the simple set of equations:
\begin{align}
\nabla^2 \mathbf{u}&=\nabla p \\
\nabla\cdot \mathbf{u}&=0\\
\partial_t \xi + \mathbf{u}\cdot\nabla \xi &=\nabla^2\xi
\end{align}%
Here, \reviewerA{$\mathbf{u}$ is the velocity, $p$ the pressure} and the mass fraction field $w$ of the nonvolatile component (e.g. hexanediol) has been replaced by the field $\xi=(w-w_\text{min})/\Ev$ centered around the surface tension minimum and normalized by the nondimensional evaporation number\cite{Diddens2021}
\begin{align}
\Ev=\frac{D_\text{vap}}{D}\frac{c_\text{sat}-c_\infty}{\rho} w_\text{avg}\,. 
\end{align}
The saturated water vapor density $c_\text{sat}$ at the liquid-gas interface and the liquid diffusivity $D$ are assumed to be constant, evaluated at the averaged mass fraction $w_\text{avg}$ in the droplet and room temperature. $c_\infty$ and $D_\text{vap}$ are the ambient vapor density and the water vapor diffusivity, respectively. For the boundary conditions, a no-slip condition at the substrate is imposed and at the liquid-gas interface, we assume no mass loss due to evaporation ($\mathbf{u}\cdot\mathbf{n}=0$), i.e. an adiabatic approximation due to the separation of time scales. The compositional change due to evaporation is evaluated in the homogeneous flat droplet limit, $\nabla \xi\cdot\mathbf{n}=\frac{2}{\pi}(1-r^2)^{-1}$ and we impose a Marangoni stress due to the nondimensionalized parabolic surface tension $\sigma=\Ma\: \xi^2/2$, i.e.
\begin{align}
\mathbf{n}\cdot(\nabla\mathbf{u}+\nabla\mathbf{u}^\text{T})\cdot \mathbf{t}=\Ma\: \xi\: \nabla_\Gamma \xi\cdot \mathbf{t}
\end{align}
with the Marangoni number $\Ma=\Ev^2 \sigma_2 R/(\mu D)$, \reviewerA{tangent $\mathbf{t}$ and normal $\mathbf{n}$} . Again, the dynamic viscosity $\mu$ is assumed to be constant. Lastly, in order to allow for stationary solutions, we fix the spatial average of $\xi$ to $\xiavg$. For a more detailed derivation of similar minimal models, the reader is referred to Refs.\cite{Diddens2021,VanGaalen2022,Rocha2025}.

To mimic the evaporation of water on this quasi-stationary model, we gradually increase the nondimensional averaged concentration of hexanediol, i.e. $\xiavg$. When $\xiavg$ is sufficiently negative, the composition field $\xi$ is negative throughout the droplet, i.e. the entire interface has a surface tension declining with increasing composition $\xi$. This gives rise to a Marangoni circulation from the contact line along the liquid-gas interface toward the apex. Upon increasing $\xiavg$, the composition attains a distribution where the region near the contact line has already passed the minimum in the surface tension (cf. \figref{fig3}(b)). This leads to a reversal of the Marangoni flow in a region of size $d_\text{rev}$ near the contact line. To predict the three-dimensional dynamics, we determined the axisymmetric quasi-stationary solutions of the minimal model and tested for azimuthal stability by perturbing the axisymmetric base state by perturbations $\propto e^{im\varphi}$, where $m$ is the mode number and $\varphi$ is the azimuthal angle. Thereby, one can extract the dominant eigenvalue and -vector for each azimuthal mode $m$ and find the critical mode $m_\text{crit}$ for which the real part of an eigenvalue crosses zero upon increasing $\xiavg$ (cf. \figref{fig3}(c) and Supplementary Movie\cite{supp}). A similar approach has been recently used to predict the symmetry-breaking of thermal Marangoni flow in nonvolatile single-component droplets on heated or cooled substrates\cite{Babor2023}. More details on our implementation of this method also can be found in Ref. \cite{Diddens2024}.
As depicted in \figref{fig3}(d), the superposition of the axisymmetric base state and the critical eigenmode shows indeed a spot formation at the rim, while the center is still axisymmetric. 
With increasing Marangoni number, more spots appear in the rim region (\figref{fig3}(e)), following the scaling $m_\text{crit}\sim \Ma^{1/2}$. The spots keep their aspect ratio, i.e. the size of the reversal region at the onset shrinks, $d_\text{rev}\sim \Ma^{-1/2}$ (inset of \figref{fig3}(e)). We also performed scans in the contact angle $\theta$. The flatter the droplet at the moment of the instability, the more spots will emerge. The prediction of the number of spots $m_\text{crit}$ as function of $\Ma$ and $\theta$ is depicted in \figref{fig3}(f).

\begin{figure}[t]
\includegraphics[width=8.6cm]{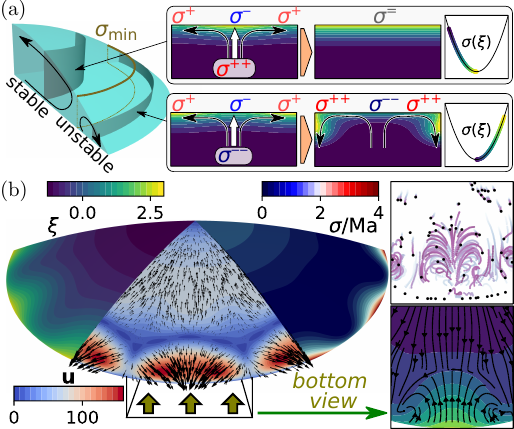}%
\caption{Mechanism of spot formation. (a) In the central region, the Marangoni instability cannot develop. Any perturbation at the interface pulls liquid with higher surface tension $\sigma^{++}$ to spots of lower surface tension $\sigma^-$ so that it equilibrates \reviewerA{(inset: cross sections at two different radii)}. In the rim region, liquid with considerably lower surface tension $\sigma^{--}$ is pulled from the bulk to the spots with $\sigma^-$, enhancing the perturbation. (b) Composition, velocity, and surface tension at the interface after onset of the instability. Once the spots are formed, strong Marangoni circulation is confined to the individual spots, allowing to quickly evaporate the remaining water by continuous mixing. As revealed in the bottom view, the numerical flow field agrees with the PTV measurements. \label{fig4}}%
\end{figure}

While the numerical analysis of the minimal model gives a quantitative prediction of the azimuthal instability that causes the breakup of the segregated rim region into spots, the physical mechanism of this process is left to be explained. It can be attributed to the Marangoni instability\cite{Schwarzenberger2014,Rocha2024marangoni}, which was first investigated by Sternling \& Scriven\cite{Sternling1959}. 
This mechanism is schematically illustrated in \figref{fig4}(a), where the stability of a tiny compositional perturbation at the interface is outlined. The disturbance introduces regions of slightly elevated and reduced surface tension, $\sigma^+$ and $\sigma^-$, respectively, which induces Marangoni flow. By virtue of continuity, this tangential Marangoni flow also induces bulk flow in the normal direction. In particular, liquid is transported from the bulk to the $\sigma^-$-spots at the interface, i.e. to spots with a lower perturbed surface tension. The stability of the interfacial perturbation now decisively depends on whether the liquid composition beneath the interface, i.e. in the normal direction, is associated with a higher or lower surface tension than at the interface.

In the central region of the droplet (top inset of \figref{fig4}(a)), the bulk liquid composition beneath the interface is associated with a higher surface tension $\sigma^{++}$ as compared to the entire surface tension at the interface, i.e. $\sigma^{++}>\sigma^{+}>\sigma^{-}$. The tangential perturbation at the interface thereby pulls up bulk liquid associated to $\sigma^{++}$ to spots of smaller perturbed surface tension $\sigma^{-}$. Thereby, the Marangoni flow ceases again, i.e. the the compositional perturbation is stable.

In the rim region (bottom inset of \figref{fig4}(a)), however, the bulk liquid is associated with a lower surface tension $\sigma^{--}$ than the surface tension at the interface, i.e. $\sigma^{--}<\sigma^{-}<\sigma^{+}$. When the Marangoni flow resulting from the interfacial perturbation pulls the bulk liquid to the perturbed $\sigma^{-}$-spots, the compositional disturbance and the resulting Marangoni flow are self-enhancing. At the same time, the liquid at the interface further evaporates while being transported to the spots of initial slightly higher surface tension $\sigma^+$, also enhancing these spots to even higher surface tension $\sigma^{++}$. 

The latter mechanism is responsible for the strong axial symmetry breaking observed e.g. in water-ethanol droplets \cite{Christy2011,Bennacer2014,Diddens2017,Rocha2024marangoni}, while the former scenario leads to regular axisymmetric Marangoni flow in e.g. water-glycerol droplets\cite{ThayyilRaju2022}.
In the scenario discussed here, with a local minimum in the surface tension curve, both cases simultaneously coexist in the same droplet: stable Marangoni flow towards the apex in the central region and unstable Marangoni flow towards the contact line in the rim region, which is separated by the line of minimal surface tension $\sigma_\text{min}$ (cf. \figref{fig4}(a)). With ongoing time, water evaporates and the line of minimal surface tension $\sigma_\text{min}$ contracts towards the center, leading to a growth of the unstable and non-axisymmetric rim region and to coalescence and coarsening by nonlinear interactions of the spots.

Lastly, we show how this azimuthal instability even enhances the evaporation-induced segregation, i.e. the intense compositional difference between the hexanediol-rich spots and the surrounding water-rich liquid. From the superposition of the axisymmetric base state and the critical eigenfunction depicted in \figref{fig4}(b), it is apparent that the spot formation induces intense Marangoni flow which is confined to each individual spot. Within each spot, this leads to quick mixing, i.e., remaining water is transported to the interface and can evaporate from each spot. Thereby, the concentration of hexanediol within each spot rises even more, giving rise to sharp fronts between the spots and the surrounding liquid (cf. \figref{fig1}). 

To summarize, we conclusively attributed the observed spot formation in evaporating water/hexanediol droplets to the evaporation-induced segregation in combination with a shallow minimum in the plateau region of the surface tension, \reviewerAB{whose presence is confirmed by multiple new accurate measurements with various different methods\cite{PRFJointSubmission}}. While it is naively unexpected that a deviation with a minimum of $~\sim \SIrange[range-phrase=-,range-units=single]{0.3}{0.7}{\milli\newton/\meter}$ in amplitude is sufficient to induce such an intense symmetry breaking, our minimal model clearly reveals that it happens even at moderate Marangoni numbers, where the number of spots scale as $m_\text{crit}\sim \sqrt{\Ma}$. Since the liquid diffusivity cubically enters the Marangoni number, a rather low diffusion coefficient can give rise to such Marangoni numbers, even if the curvature $\sigma_2$ of the surface tension curve is tiny. 

This work gives new insights in order to fully control the flow in evaporating droplets, with utmost relevance for applications like inkjet printing, surface coating and particle patterning. 
In a broader context, related phenomena, where non-monotonic material properties lead to unexpected phenomena, are the importance of non-monotonic viscosity in viscous fingering\cite{Haudin2016} and the impact of the non-monotonic density of water on the scallop formation of melting ice\cite{Weady2022}.

\begin{acknowledgments}
This work was supported by an Industrial Partnership Programme of the Netherlands Organisation for Scientific Research (NWO) \& High Tech Systems and Materials (HTSM), co-financed by Canon Production Printing Netherlands B.V. and the University of Twente. 
\end{acknowledgments}

\bibliography{manuscript}

\end{document}